\documentclass[twoside,twocolumn,9pt]{article}
\usepackage{extsizes}
\usepackage[super,sort&compress,comma]{natbib} 
\usepackage[version=3]{mhchem}
\usepackage[left=1.5cm, right=1.5cm, top=1.785cm, bottom=2.0cm]{geometry}
\usepackage{balance}
\usepackage{mathptmx}
\usepackage{amsmath}
\usepackage{amssymb}
\usepackage{sectsty}
\usepackage{graphicx} 
\usepackage{lastpage}
\usepackage[format=plain,justification=justified,singlelinecheck=false,font={stretch=1.125,small,sf},labelfont=bf,labelsep=space]{caption}
\usepackage{float}
\usepackage{fancyhdr}
\usepackage{fnpos}
\usepackage[english]{babel}
\addto{\captionsenglish}{%
  
}
\usepackage{array}
\usepackage{droidsans}
\usepackage{charter}
\usepackage[T1]{fontenc}
\usepackage[usenames,dvipsnames]{xcolor}
\usepackage{setspace}
\usepackage[compact]{titlesec}
\usepackage{hyperref}
\usepackage{epstopdf}%This line makes .eps figures into .pdf - please comment out if not required.

\definecolor{cream}{RGB}{222,217,201}
\definecolor{jlblue}{rgb}{0.0,0.6056031611752245,0.9786801175696073}
\definecolor{jlorange}{rgb}{0.8888735002725198,0.43564919034818994,0.2781229361419438}
\definecolor{jlgreen}{rgb}{0.2422242978521988,0.6432750931576305,0.3044486515341153}
\definecolor{jlviolet}{rgb}{0.7644401754934356,0.4441117794687767,0.8242975359232758}

\begin{document}

\pagestyle{fancy}
\thispagestyle{plain}
\fancypagestyle{plain}{
%%%HEADER%%%
\renewcommand{\headrulewidth}{0pt}
}
%%%END OF HEADER%%%

%%%PAGE SETUP - Please do not change any commands within this section%%%
\makeFNbottom
\makeatletter
\renewcommand\LARGE{\@setfontsize\LARGE{15pt}{17}}
\renewcommand\Large{\@setfontsize\Large{12pt}{14}}
\renewcommand\large{\@setfontsize\large{10pt}{12}}
\renewcommand\footnotesize{\@setfontsize\footnotesize{7pt}{10}}
\makeatother

\renewcommand{\thefootnote}{\fnsymbol{footnote}}
\renewcommand\footnoterule{\vspace*{1pt}% 
\color{cream}\hrule width 3.5in height 0.4pt \color{black}\vspace*{5pt}} 
\setcounter{secnumdepth}{5}

\makeatletter 
\renewcommand\@biblabel[1]{#1}            
\renewcommand\@makefntext[1]% 
{\noindent\makebox[0pt][r]{\@thefnmark\,}#1}
\makeatother 
\renewcommand{\figurename}{\small{Fig.}~}
\sectionfont{\sffamily\Large}
\subsectionfont{\normalsize}
\subsubsectionfont{\bf}
\setstretch{1.125} %In particular, please do not alter this line.
\setlength{\skip\footins}{0.8cm}
\setlength{\footnotesep}{0.25cm}
\setlength{\jot}{10pt}
\titlespacing*{\section}{0pt}{4pt}{4pt}
\titlespacing*{\subsection}{0pt}{15pt}{1pt}
%%%END OF PAGE SETUP%%%

%%%FOOTER%%%
\fancyfoot{}
\fancyfoot[LO,RE]{\vspace{-7.1pt}\includegraphics[height=9pt]{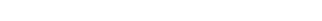}}
\fancyfoot[CO]{\vspace{-7.1pt}\hspace{13.2cm}\includegraphics{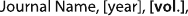}}
\fancyfoot[CE]{\vspace{-7.2pt}\hspace{-14.2cm}\includegraphics{head_foot/RF}}
\fancyfoot[RO]{\footnotesize{\sffamily{1--\pageref{LastPage} ~\textbar  \hspace{2pt}\thepage}}}
\fancyfoot[LE]{\footnotesize{\sffamily{\thepage~\textbar\hspace{3.45cm} 1--\pageref{LastPage}}}}
\fancyhead{}
\renewcommand{\headrulewidth}{0pt} 
\renewcommand{\footrulewidth}{0pt}
\setlength{\arrayrulewidth}{1pt}
\setlength{\columnsep}{6.5mm}
\setlength\bibsep{1pt}
%%%END OF FOOTER%%%

%%%FIGURE SETUP - please do not change any commands within this section%%%
\makeatletter 
\newlength{\figrulesep} 
\setlength{\figrulesep}{0.5\textfloatsep} 

\newcommand{\topfigrule}{\vspace*{-1pt}% 
\noindent{\color{cream}\rule[-\figrulesep]{\columnwidth}{1.5pt}} }

\newcommand{\botfigrule}{\vspace*{-2pt}% 
\noindent{\color{cream}\rule[\figrulesep]{\columnwidth}{1.5pt}} }

\newcommand{\dblfigrule}{\vspace*{-1pt}% 
\noindent{\color{cream}\rule[-\figrulesep]{\textwidth}{1.5pt}} }

\makeatother
%%%END OF FIGURE SETUP%%%

%%%TITLE, AUTHORS AND ABSTRACT%%%
\twocolumn[
  \begin{@twocolumnfalse}
{\includegraphics[height=30pt]{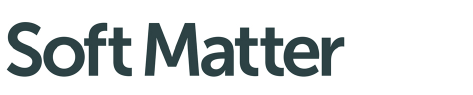}\hfill\raisebox{0pt}[0pt][0pt]{\includegraphics[height=55pt]{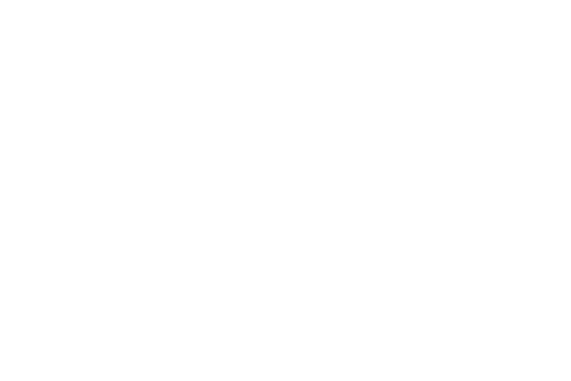}}\\[1ex]
\includegraphics[width=18.5cm]{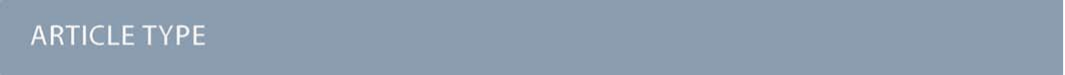}}\par
\vspace{1em}
\sffamily
\begin{tabular}{m{4.5cm} p{13.5cm} }

\includegraphics{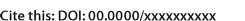} & \noindent\LARGE{\textbf{Instabilities of ring-rivulets: Impact of substrate wettability}} \\%Article title goes here instead of the text "This is the title"
\vspace{0.3cm} & \vspace{0.3cm} \\

 & \noindent\large{Stefan Zitz,\textit{$^{a\ast}$} Andrea Scagliarini,\textit{$^{b,c}$} Johan Roenby\textit{$^{a\dagger}$}} \\%Author names go here instead of "Full name", etc.

\includegraphics{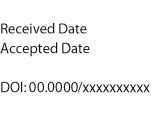} & \noindent\normalsize{
\noindent Rivulets and droplets are naturally appearing shapes when small amounts of liquid are deposited on a partially wettable substrate.
Here we study, by means of numerical simulations, the dewetting dynamics of a ring-rivulet on substrates with various contact angles and wettability patterns.
In particular, we consider, beyond the homogeous case, an annular band of lower contact angle as compared to the background and a constant radial gradient of contact angle, pointing either inward or outward from the centre. 
We show that by tuning the parameters characterizing the patterns, it is possible to control 
not only the stability of the rivulet, i.e. its breakup/collapse dynamics and 
the associated time scales, but also the dewetting morphology, in terms of number and position of the formed droplets.
%Thus making it possible to enhance the probability for a Rayleigh-Plateau type %break up into multiple droplets or to accelerate the capillary retraction that %stems from the curvature difference of the ring and leads to a single droplet.
}

\end{tabular}

 \end{@twocolumnfalse} \vspace{0.6cm}

]
%%%END OF TITLE, AUTHORS AND ABSTRACT%%%

%%%FONT SETUP - please do not change any commands within this section
\renewcommand*\rmdefault{bch}\normalfont\upshape
\rmfamily
\section*{}
\vspace{-1cm}

%%%FOOTNOTES%%%

\footnotetext{\textit{$^{a}$~IMFUFA, Department of Science and Environment, Roskilde University, Postbox 260, 4000 Roskilde, DK.}}
\footnotetext{\textit{$^{\ast}$~current E-mail: stefan.zitz@rcpe.at}}
\footnotetext{\textit{$^{\dagger}$~Tel: +45 2993 1923; E-mail: johan@ruc.dk}}
\footnotetext{\textit{$^{b}$~Institute for Applied Mathematics "M. Picone" (IAC), Consiglio Nazionale delle Ricerche (CNR), Via dei Taurini 19, 00185 Rome, Italy, E-mail: andrea.scagliarini@cnr.it}}
\footnotetext{\textit{$^{c}$~INFN, sezione Roma ``Tor Vergata'', via della Ricerca Scientifica 1, 00133 Rome, Italy}}

%%%END OF FOOTNOTES%%%

%%%MAIN TEXT%%%%
\section{Introduction}
\label{sec:intro}
Thin liquid films and droplets are widespread in our every day life and play a crucial role in a host of natural and technological applications, from painting and coating to lab-on-a-chip devices to biofluidics~\cite{degennesCapillarityWettingPhenomena2004, ronsinPhaseFieldSimulationsMorphology2022,fockeLabonaFoilMicrofluidicsThin2010}.
Understanding their dynamics and controlling their stability is, therefore, a central problem for applied research in process engineering and nanotechnology~\cite{singhInkjetPrintingProcess2010, quereFluidCoatingFiber1999, utadaDrippingJettingDrops2007}, but also poses fundamental questions lying at the crossroads between fluid dynamics and chemical physics~\cite{oronLongscaleEvolutionThin1997, beckerComplexDewettingScenarios2003, thielePatternedDepositionMoving2014, wilczekSlidingDropsEnsemble2017, peschkaSignaturesSlipDewetting2019}.
Dewetting induced by intrinsic instabilities of the film and/or impurities on the substrate surface, for instance, can undermine the effectiveness of a coating process~\cite{bonnWettingSpreading2009, chenWrinklingInstabilitiesPolymer2012}. 
On the other hand, breakup of deposited structures such as rivulets is exploited in the generation of droplets {\it on demand}~\cite{nguyenCompetitionCollapseBreakup2012, PhysRevLett.133.214003}.
All these phenomena involve inherently multiscale problems, that span from the molecular motion at the three phase contact line to the nano-/mircoscale thickness of the film up to the macroscopic area the coating covers, thus posing non trivial computational challenges.
The dewetting of a fluid rivulet deposited on a substrate recalls the classical fluid dynamic problem of a filament breakup, driven by the Rayleigh-Plateau instability, with the additional complexity of the fluid-solid physico-chemical interactions~\cite{diezBreakupFluidRivulets2009, diezStabilityFinitelengthRivulet2009, diezInstabilityTransverseLiquid2012}.
Recently,  a number of experimental and theoretical/numerical studies have focused on ring-shaped rivulets, that were shown to be useful precursors of droplet patterns with a circular symmetry~\cite{nguyenCompetitionCollapseBreakup2012, gonzalezStabilityLiquidRing2013, wuCompetingLiquidPhase2011, edwardsControllingBreakupToroidal2021}. 
Here, at difference with the simpler straight rivulet case, the dewetting dynamics depends also on the non-uniform curvature of the rivulet ringed shape, and in particular on the different curvatures of the two contact lines.
Self- and direct assembly of nanomaterials from liquid nanostructures has been one of the driving forces for the study of Nguyen et al.~\cite{nguyenCompetitionCollapseBreakup2012} and earlier studies of Wu et al.~\cite{wuBreakupPatternedNanoscale2010} where they showed that liquid-metal rings are suited to form arrays of droplets.
Diez et al.~\cite{diezBreakupFluidRivulets2009, diezStabilityFinitelengthRivulet2009} laid the theoretical foundation for the stability of a straight rivulet in their work using linear stability analysis (LSA) and numerical simulations. 
Later, Gonz{\'a}lez et al.~\cite{gonzalezStabilityLiquidRing2013} extended these results to ring rivulets, determining the characteristic time scales for collapse and breakup and showing that the main control parameter discriminating between the two instability routes is the rivulet aspect-ratio, namely the ratio of its width over the its radius.
They also provided predictions on the expected number of formed droplets as dictated by the most unstable wavelength.
An interesting question that may naturally arise is whether and to which extent it is possible to further control the fate of ring rivulets and the consequent dewetting morphologies by properly treating the substrate to exploit wettability patterns. 
With recent developments in surface chemistry and the emerging technology of switchable substrates local precise wettability gradients are now readily attainable~\cite{xinReversiblySwitchableWettability2010, stuartEmergingApplicationsStimuliresponsive2010,chenThermalresponsiveHydrogelSurface2010, ichimuraLightDrivenMotionLiquids2000, mugeleElectrowettingConvenientWay2005, edwardsControllingBreakupToroidal2021}.
While a number of studies have focused on thin film dewetting and droplet transport on patterned substrates~\cite{liuActuatingWaterDroplets2015,grawitterSteeringDropletsSubstrates2021, zitzControllingDewettingMorphologies2023}, the case of ring-shaped rivulets remained so far almost unexplored.
A relevant exception is represented by the work of Edwards et al.~\cite{edwardsControllingBreakupToroidal2021}, who have studied numerically and experimentally liquid rings deposited on a substrate Where the contact angle was controlled by electrowetting. 
They showed that different electric potentials can be used not only to control the number of droplets after breakup but also to fully reverse the process.

In the present paper, we perform a systematic study of the effect of the ring rivulet initial geometry (width over radius ratio) and of two types of wettability patterns in the selection of route towards either retraction and collapse to a single droplet or towards the breakup into multiple droplets. 
We show that by depositing the liquid ring onto an annular region of lower contact angle (with respect to the background substrate) one can remove the collapse mode. 
Moreover, the contact angle contrast turns out to serve as a control parameter determining together with the dimensionless initial ring width the number of droplets in the final stationary equilibrium state.
By introducing, instead, a radially symmetric linear contact angle profile, coaxial with the ring and pointing either inwards or outwards, we find that it is possible, in the former case, to control the retraction speed while steering the number of metastable droplets whereas, in the latter case, to stabilize the ring rivulet against collapse.
The outline of the paper is as follows: In the next section, Sec.~\ref{sec:method} we introduce the method we use to run numerical experiments.
We then present our results in Sec.~\ref{sec:results}, starting with a comparison with the literature and then present the impact of the wettability patterns.
In the last section, Sec.~\ref{sec:conclu} we give a short summary, highlighting important results and conclude with an outlook of possible research applications.

\section{Simulation method}
\label{sec:method}
We perform numerical simulations of the thin film equation (TFE),  
\begin{equation}\label{eq:thinsolve}
     \partial_t h(\mathbf{x},t) = \nabla\cdot\left(M_{\delta}(h)\nabla p\right),
\end{equation}
where $\mathbf{x} = (x,y)$ and $\nabla = (\partial_x, \partial_y)$, by means of a recently developed lattice Boltzmann method with an associated solver called Swalbe~\cite{zitzLatticeBoltzmannMethod2019, zitzLatticeBoltzmannSimulations2021, zitzSwalbeJlLattice2022, zitzControllingDewettingMorphologies2023}. 
The mobility function $M_{\delta}(h) = \frac{h^2}{\mu\alpha_{\delta}(h)}$ with 
\begin{equation}\label{eq:alphafric}
    \alpha_{\delta}(h) = \frac{6h}{(2 h^2 + 6 \delta h + 3 \delta^2)},
\end{equation}
which for the no-slip boundary condition $(\delta \rightarrow 0)$ becomes $M_{0}(h) = h^3/3\mu$, where $\delta$ is an effective slip length and $\mu$ is the dynamic viscosity, both values can be found in App.~\ref{app:numerics}.
We like to point out, however, that the slip length value is within the weak/intermediate slip regime~\cite{peschkaSignaturesSlipDewetting2019,fetzerQuantifyingHydrodynamicSlip2007, munchLubricationModelsSmall2005} and has been used in previous work~\cite{zitzControllingDewettingMorphologies2023}.
The pressure $p$ in Eq.~(\ref{eq:thinsolve}) is given by,
\begin{equation}\label{eq:filmpressure}
    p = - \gamma\nabla^2 h -\Pi(h),
\end{equation}
with $\Delta h$ being the 2D Laplacian of the liquid-gas interface and $\Pi(h)$ is a so-called disjoining pressure~\cite{schwartzSimulationDropletMotion1998, crasterDynamicsStabilityThin2009, nguyenCompetitionCollapseBreakup2012, gonzalezStabilityLiquidRing2013}
\begin{equation}\label{eq:disjoinpressure}
    \Pi(h,\theta) = \frac{2\gamma}{h_{\ast}}[1-\cos\theta(\mathbf{x})]\left[\left(\frac{h_*}{h}\right)^3 -\left(\frac{h_*}{h}\right)^2\right],
\end{equation}
where $\gamma$ is the surface tension, $h_{\ast}$ is a precursor thickness, see App.~\ref{app:numerics}, at which $\Pi(h_{\ast}, \theta) = 0$ and $\theta$ is an equilibrium contact angle.
By allowing spatial variation of $\theta$ in Eq.~(\ref{eq:disjoinpressure}) we have an effective model for a patterned substrate, see e.g., Refs~\cite{zitzLatticeBoltzmannSimulations2021, zitzControllingDewettingMorphologies2023}. 
The contact angle, in agreement with the lubrication approximation~\cite{oronLongscaleEvolutionThin1997, crasterDynamicsStabilityThin2009}, is set within the bounds $[\pi/18, 2\pi/9]$, except for the banded pattern, see Sec.~\ref{subsubsec:banded}. 

\subsection{Initial conditions}
\begin{figure}
\centering
  \includegraphics[width=0.45\textwidth]{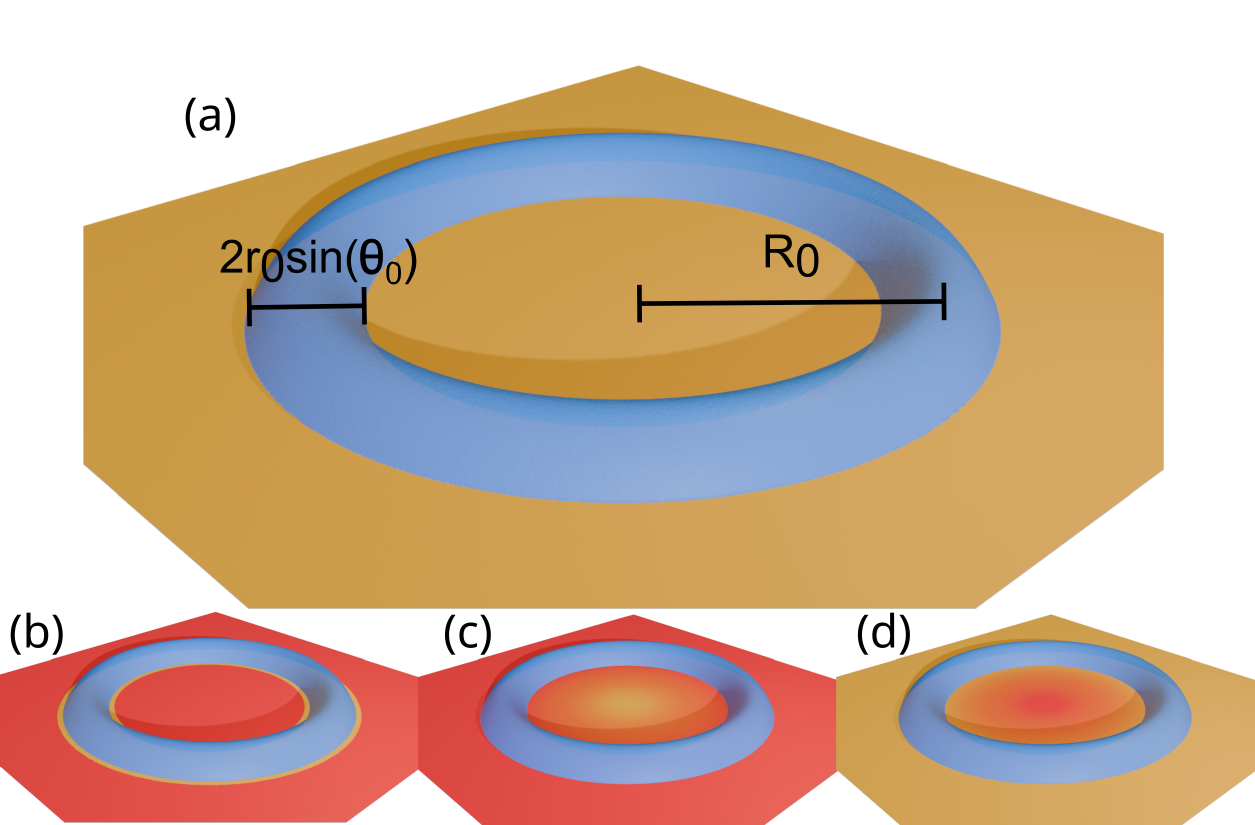}
  \caption{Schematic setup of our initial conditions. In (a) we show a render of the initial fluid state and display $R_0$ and $r_0$. 
  Below in panels (b)-(d) we show different wettability patterns with yellow being more wettable than red.
  While (b) depicts the band pattern, (c) and (d) show radial gradient wettabilities.} 
  \label{fig:ringschema}
\end{figure}

We initialise the thickness profile $h_0(\mathbf{x})=h(\mathbf{x}, t=0)$ by first imposing the shape of a toroidal cap, with radial symmetry along the $z$-axis, centred in the origin of the coordinate system and with major an minor radii $R_0$ and $r_0$, whose equation in polar coordinates $(\xi, \phi)$ (with $x = \xi \cos(\phi)$, $y=\xi \sin(\phi)$ reads,
\begin{equation}\label{eq:torus}
h_0(\mathbf{x})=\left(\sqrt{r_0^2 - \left(R_0-\xi\right)^2} - r_0\cos \theta_0 + h_{\ast}\right)
\end{equation}
for $|R_0-\xi|<r_0 \sin \theta_0$ (and $h_0(\mathbf{x})=h_{\ast}$, otherwise); then, we let it relax to the actual equilibrium shape~\cite{diezBreakupFluidRivulets2009} which we slightly perturb with a Gaussian noise with zero mean and variance $10^{-4}r_0^2 \sin^2\theta_0$. 

\section{Results} \label{sec:results}

\subsection{Dewetting on uniform substrates}\label{subsec:drop-counting}
\begin{figure}
    \centering
    \includegraphics[width = 0.45\textwidth]{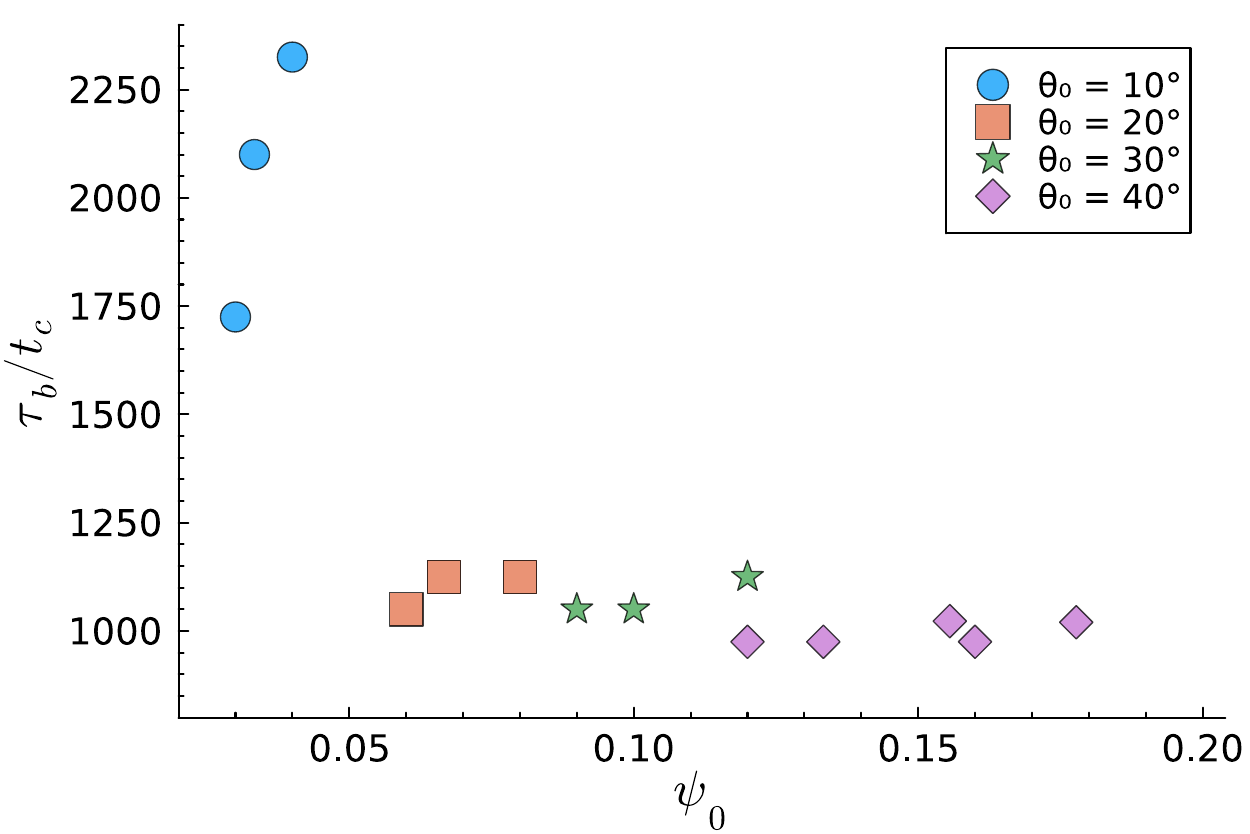}
    \caption{Rivulet breakup times (in units of $t_c = \mu r_0/\gamma$) for various substrate contact angle as a function of the initial aspect-ratio $\psi_0$.}
    \label{fig:breakuptimes}
\end{figure}

We start our analysis of the dewetting dynamics of a ring-rivulet considering substrates with uniform wettability (i.e. constant contact angle $\theta(\mathbf{x}) = \theta_0$, see Fig.~\ref{fig:ringschema}(a)). 
In order to discriminate between the two available dewetting paths, we look at the final number of droplets $n_d$, into which the ring-rivulet forms, as a function of $\psi_0 \equiv 2r_0\sin(\theta_0)/R_0$.
\begin{figure}
    \centering
    \includegraphics[width=0.45\textwidth]{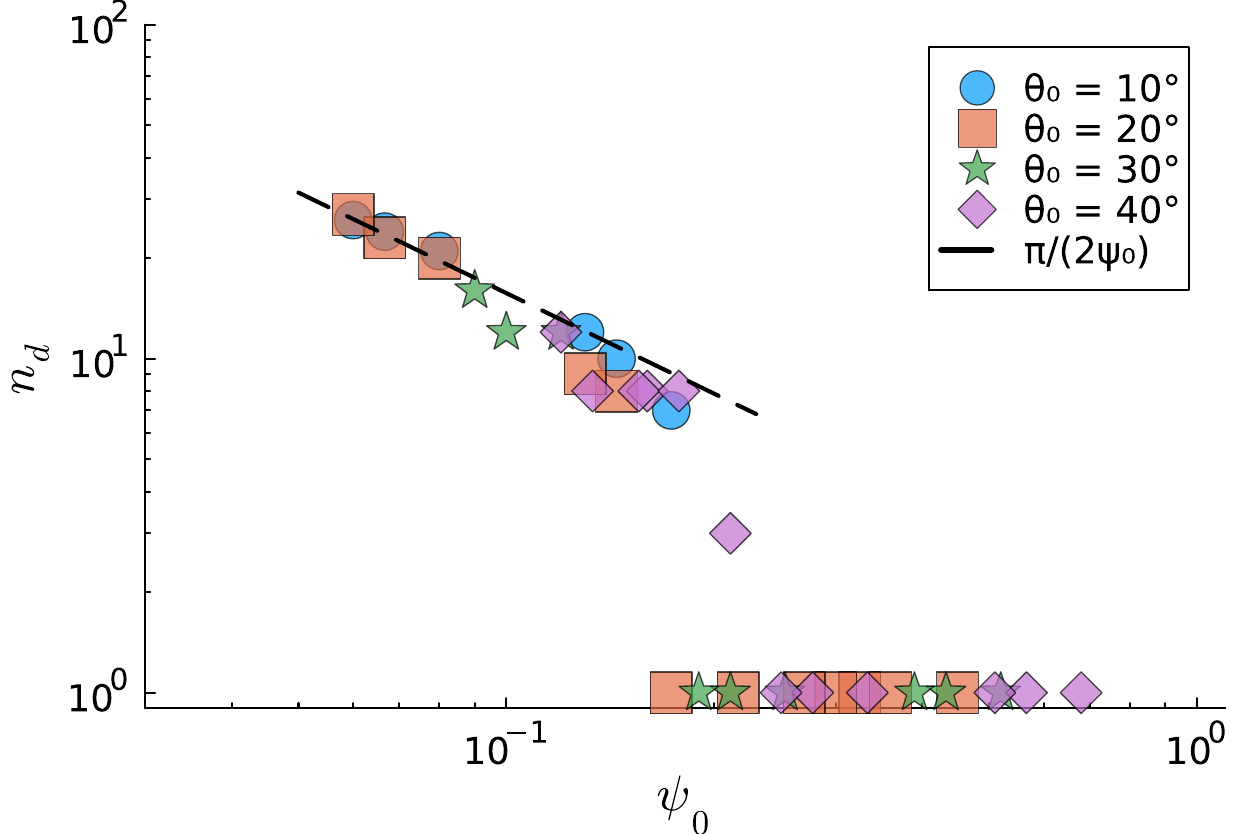}   
    \caption{Droplet number $n_d$ vs aspect ratio $\psi_0$ on uniform substrates for various contact angles $\theta_0$. 
    The solid line depicts the approximated LSA prediction, Eq.~(\ref{eq:maxDrops})~\cite{gonzalezStabilityLiquidRing2013}.}
    \label{fig:max_drops}
\end{figure}
Fig.~\ref{fig:max_drops} shows, $n_d$ vs. $\psi_0$ for various contact angles. 
For small aspect-ratio, $\psi_0 <0.2$, $n_d>1$, indicating that the rivulet undergoes a breakup, whereas for $\psi_0>0.2$ we get $n_d=1$, indicating the retractive collapse to a single central droplet.
When breakup takes place, moreover, we observe that $n_d$ decreases with $\psi_0$ in agreement with the approximated formula 
\begin{equation}\label{eq:maxDrops}
    n_d \approx \frac{\pi}{2\psi_0},
\end{equation}
(solid line in Fig.~\ref{fig:max_drops}) derived from linear stability analysis (LSA) assuming that the most unstable mode determines the number of droplets~\cite{gonzalezStabilityLiquidRing2013}.
The agreement is particularly good for small contact angles, as expected since the theory leading to (\ref{eq:maxDrops}) does not account for the disjoinin pressure~\cite{gonzalezStabilityLiquidRing2013}. 
Interestingly, as the contact angle increases, the number of droplets becomes almost independent of $\psi_0$, suggesting that the breakup process is mainly driven by the reduced wettability of the substrate.
Taking $\psi_0 \approx 0.2$, at which $n_d$ drops to one, then, as the value discriminating between breakup and collapse, we now focus on the characteristic times of both processes.
Hereafter times are made dimensionless by a characteristic capillary time $t_c$, defined as $t_c = \mu r_0/\gamma$.
In Fig.~\ref{fig:breakuptimes} we report the breakup times as a function of $\psi_0$.
The breakup time, $\tau_b$, is defined as the earliest instant of time at which the line $h(R(t),\phi,t)$ "touches" the substrate, namely
\begin{equation}\label{eq:breakuptime}
\tau_b = \min_t \{t | h(\xi=R(t),\phi,t) = h_{\ast}\}.
\end{equation}
For the lowest contact angle, $\tau_b$ grows with $\psi_0$. 
Such behaviour was indeed predicted by the theoretical approach of Gonz\'alez et al.~\cite{gonzalezStabilityLiquidRing2013}, assuming that the growth rate of the instability in the linear regime determines also the time scales of breakup. 
For larger contact angle ($\theta_0 > 10^{\circ}$), i.e. when disjoining pressure effects start to be relevant, instead, the breakup times tend to become almost independent of $\psi_0$, suggesting that the time scales are dictated essentially by the substrate wettability.\\
\begin{figure}
    \centering
    \includegraphics[width = 0.45\textwidth]{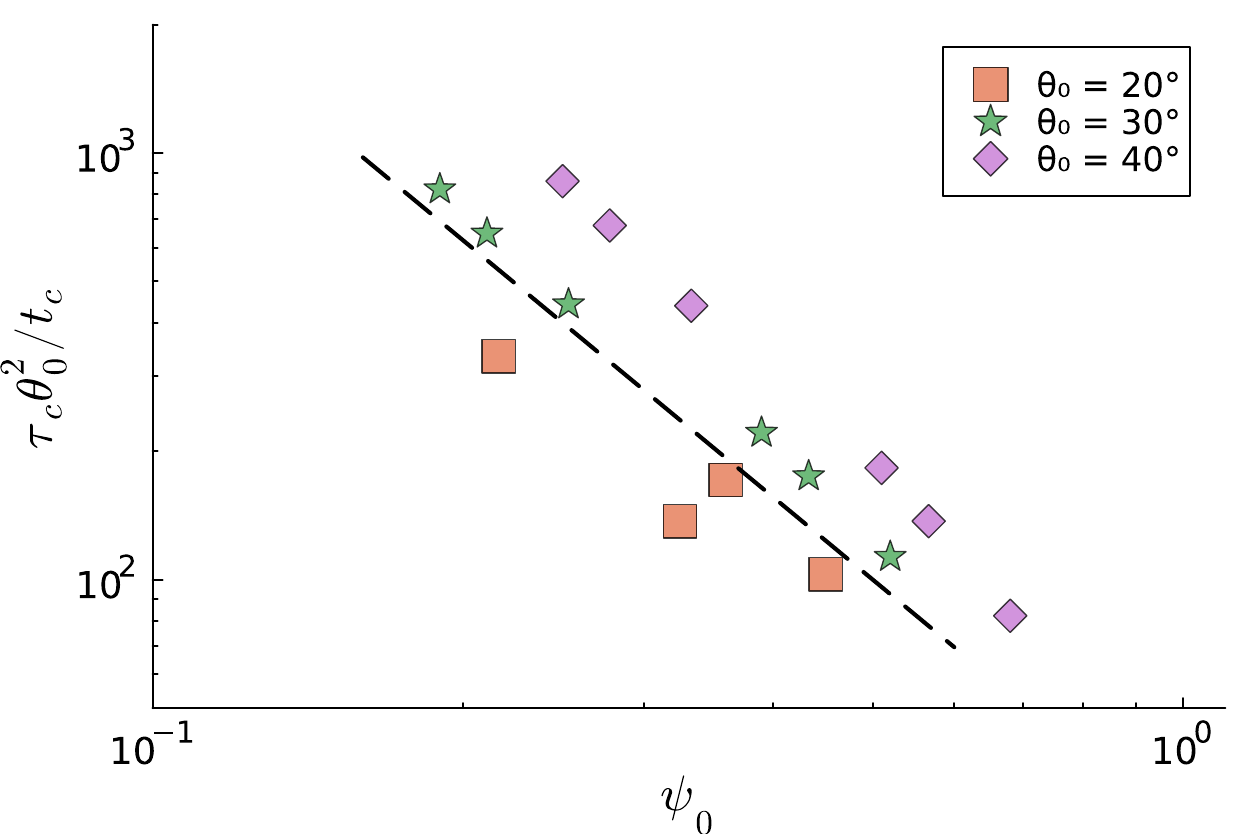}
    \caption{Log-log plot of the rivulet collapse times (in units of $t_c = \mu r_0/\gamma$) for various substrate contact angle as a function of the initial aspect-ratio $\psi_0$; the solid line indicates the power law $\tau_c \theta_0^2/t_c \sim \psi_0^{-2}$ (see Eq.(\ref{eq:modeltauc}) and related discussion for further details).}
    \label{fig:collapsetimes}
\end{figure}
Fig.~\ref{fig:collapsetimes} shows the time $\tau_c$ taken by the liquid to fully wet the hole delimited by the ring rivulet which collapses into a single droplet (hence the name "collapse time"), i.e. 
\begin{equation}\label{eq:collapsetime}
\tau_c = \min_t \{t | h(\xi=0,\phi,t) > h_{\ast}\}.
\end{equation}
Upon rescaling by $t_c$ and multiplying by $\theta_0^2$ we observe that the collapse times decay with the initial aspect-ratio as $\tau_c \theta_0^2/t_c \sim \psi_0^{-2}$; such behaviour is explained in what follows.
The collapse time can be seen as the time the point (in the radial coordinate) $\xi_1(t) = R(t) - w(t) \approx R(t)$\footnote{This approximation holds since $R \gg w$ most of the time during the retraction process} needs to reach the origin $\xi = 0$. 
We prove in the Appendix (see section \ref{sec:derivation}) that, under certain assumptions, the following differential equation for $R(t)$ holds:
\begin{equation}\label{eq:modelC4}
-\frac{d R^2}{dt} \approx 3 \, \theta_0^4 \, \frac{r_0^2}{t_c}.
\end{equation}
The latter equation can be integrated to $R_0^2 - R^2(t) \approx 3 \theta_0^4 r_0^2 t/t_c$.
The collapse time, then, is such that $R^2(\tau_c) \approx 0$, whence $\tau_c  \approx R_0^2 t_c/(3 r_0^2 \theta_0^4)$, which can be recast into (recall that $\psi_0 \approx 2 r_0 \theta_0/R_0$ for small $\theta$) 
\begin{equation}\label{eq:modeltauc}
\frac{\tau_c}{t_c} \sim  (\theta_0 \psi_0)^{-2},
\end{equation}
shown with a solid line in Fig.~\ref{fig:collapsetimes}.

\subsection{Wettability patterns}\label{subsec:wettability}
We now focus on the rivulet stability and dewetting morphology on substrates with space-varying contact angle (i.e. {\it patterned substrates}).
As anticipated in the introduction, we consider two different wettability patterns: i) an annular band 
\begin{equation}\label{eq:theta_band}
    \theta(\xi) =\begin{cases}
        \theta_a \quad \text{for}~|R_0-\xi| < r_0\sin(\theta_0) \\
        \theta_b\quad \text{otherwise}
    \end{cases},
\end{equation}
ii) an axially symmetric linear contact angle profile
\begin{equation}\label{eq:theta_grad}
    \theta(\xi) =\begin{cases}
        \frac{(\theta_{b}-\theta_{a})}{R_0} \xi + \theta_{a} \quad \text{for}~\xi \leq R_0 \\
        \theta_b \quad \text{otherwise}
    \end{cases},
\end{equation}
where either $\theta_{b} > \theta_{a}$ (outward pointing gradient) or $\theta_b < \theta_a$ (inward pointing gradient).
The different contact angle patterns are then used in Eq.~(\ref{eq:disjoinpressure}) which, in turn, enters in Eq.~(\ref{eq:thinsolve}) through the total pressure, Eq.~(\ref{eq:filmpressure}).
The idea behind such choices is that the former serves as an effective boundary removing the collapse mode, whereas the latter illuminates the force balance between wetting and retraction of the ring-rivulet, see Eqs.~(\ref{eq:theta_band}-\ref{eq:theta_grad}).

\subsubsection{Annular band}\label{subsubsec:banded}
\begin{figure}
    \centering
    \includegraphics[width=0.45\textwidth]{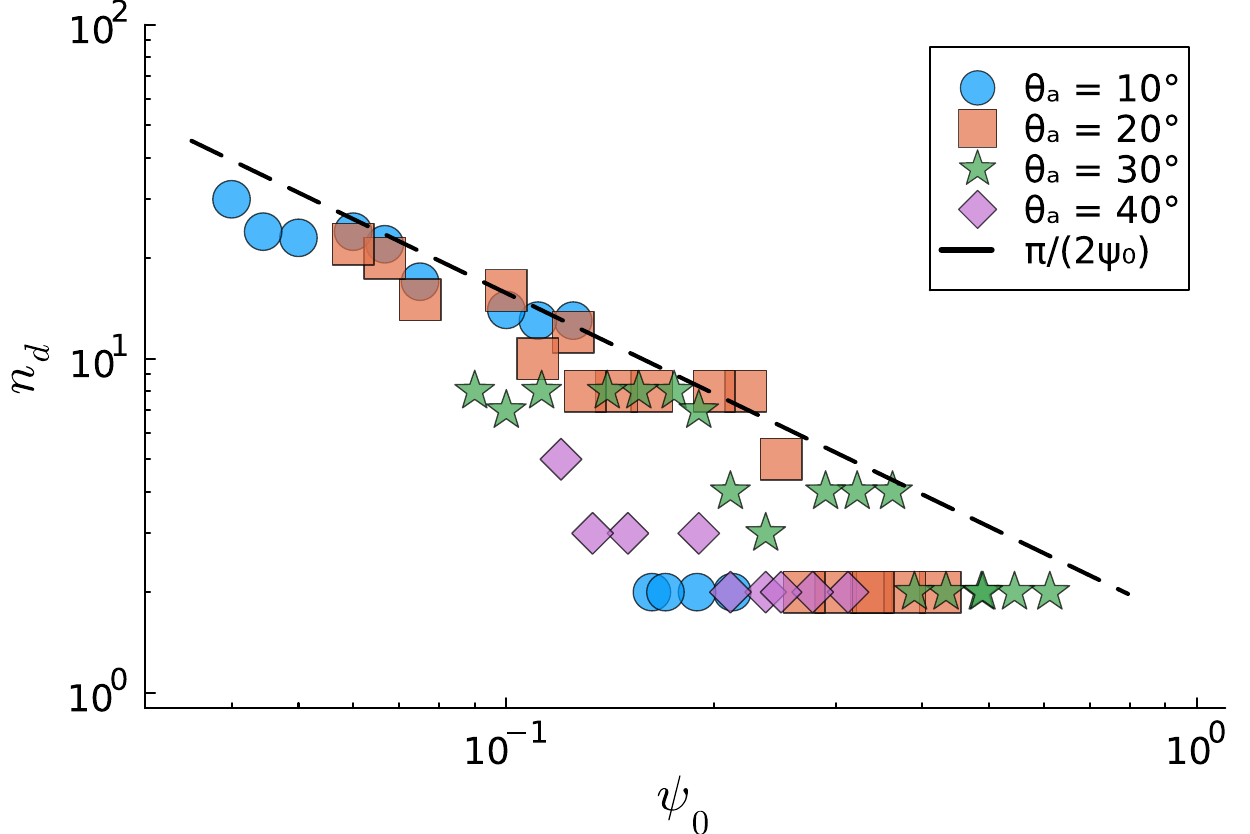}    
    \caption{Droplet number $n_d$ vs aspect ratio $\psi_0$ on substrates patterned with the annular band, Eq.~(\ref{eq:theta_band}), for various values of $\theta_a$ and $\theta_b = 60^{\circ}$. 
    The solid line depicts the approximated LSA prediction, Eq.~(\ref{eq:maxDrops})~\cite{gonzalezStabilityLiquidRing2013}.}
    \label{fig:max_drops_band}
\end{figure}

This pattern, Eq.~(\ref{eq:theta_band}), is realized by using $\theta_a \equiv \theta_0 \in [10^{\circ}, 20^{\circ}, 30^{\circ}, 40^{\circ}]$ and $\theta_b = 60^{\circ}$, thus the ring rivulet is confined on the annular band and the inner hole essentially represents an effective energetic barrier for the collapse to take place.
This setup, somehow, echoes the experiments by Edwards et al.~\cite{edwardsControllingBreakupToroidal2021}, where they managed to remove the collapse mode by means of a suitable electrowetting-based manipulation of the substrate.
This can be seen clearly in Fig.~\ref{fig:max_drops_band}, where we report the number of droplets formed upon dewetting, noticing that $n_d>1$ for every $\psi_0$ and contact angle. 
At odds with the uniform case (see Fig.~\ref{fig:max_drops}), moreover, the data correlates less with the LSA prediction $n_d = \pi/(2\psi_0)$, except for very small contact angle. 
For $\psi \gtrsim 0.1$, $n_d$ displays a sort of step-wise dependence on $\psi_0$. 
It appears that, by disentangling the radial dynamics driven by contact angle curvature imbalance, the pattern allows for the formation of a larger number of droplets even for relatively larger values of the aspect-ratio (we observe still $n_d \sim O(10)$ up to $\psi_0 \approx 0.2$).
\begin{figure}
    \centering
    \includegraphics[width=0.48\textwidth]{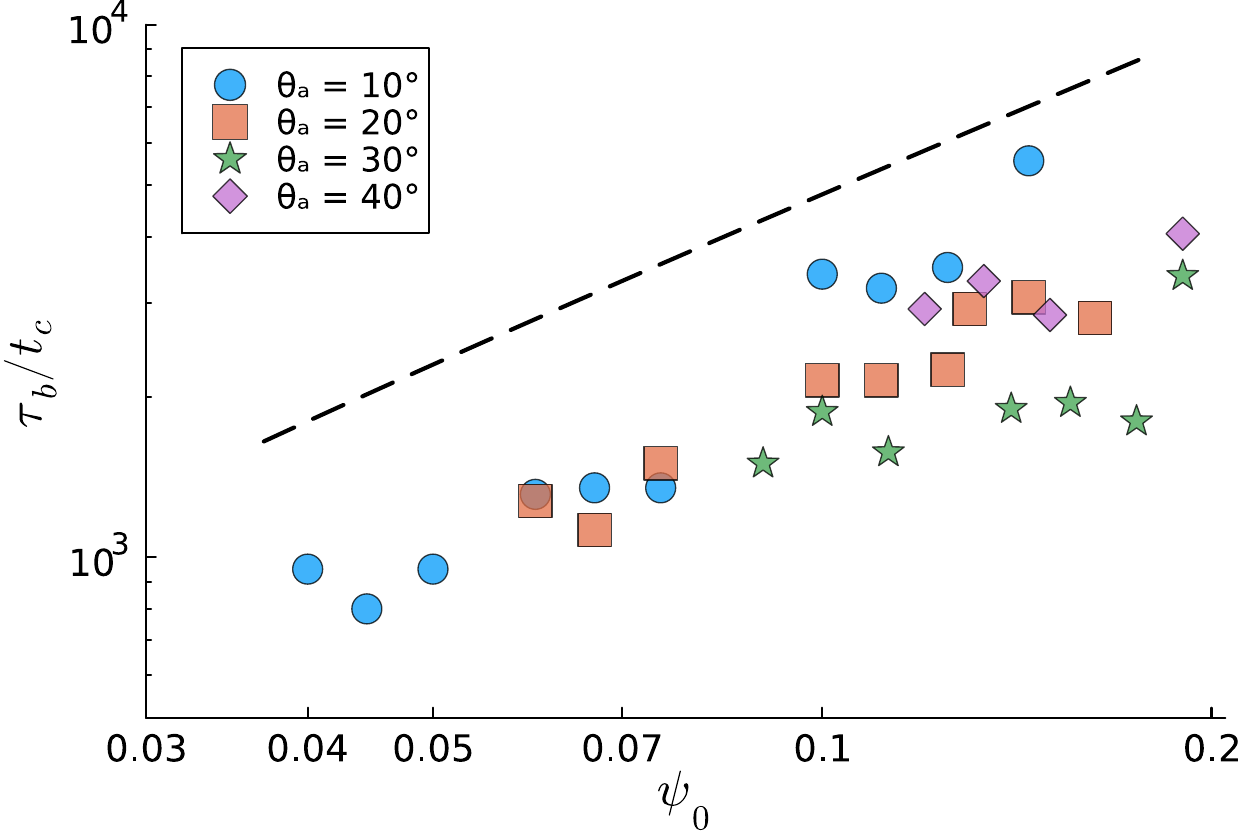}
    \caption{Log-log plot of the ring rivulet breakup times as a function of the initial aspect-ratio, for the annular band pattern, Eq.~\ref{eq:theta_band}.
    Different colors depict different value of the contact angle contrasts $\theta_a$ ($\theta_b = 60^{\circ}$ is kept fixed), see Eq.~(\ref{eq:theta_band}).
    The linear function of $\psi_0$ is reported with a dashed line as a guide for the eye.
    }
    \label{fig:bandBreakupT}
\end{figure}
In Fig.~\ref{fig:bandBreakupT}, we show the dependence of the breakup times on $\psi_0$ for different contact angle contrasts. 
For large $\theta_a$, i.e. small contact angle contrast, hence when the constraint on the band is less effective, the breakup time grows with $\psi_0$, in agreement with the prediction of the LSA~\cite{gonzalezStabilityLiquidRing2013}. 
As $\theta_a$ decreases, though, the confining effect decouples radial dynamics and dewetting and $\tau_b$ tends to become almost independent of $\psi_0$, but determined, instead, by the local contact angle. 
The breakup times are larger than in the uniform substrate case (cf. Fig~\ref{fig:breakuptimes}): so, overall the annular band pattern makes the ring rivulet slightly more stable against rupture.

\subsubsection{Linear radial profile}\label{subsubsec:linwettgrad}
\begin{figure}
    \centering
    \includegraphics[width=0.48\textwidth]{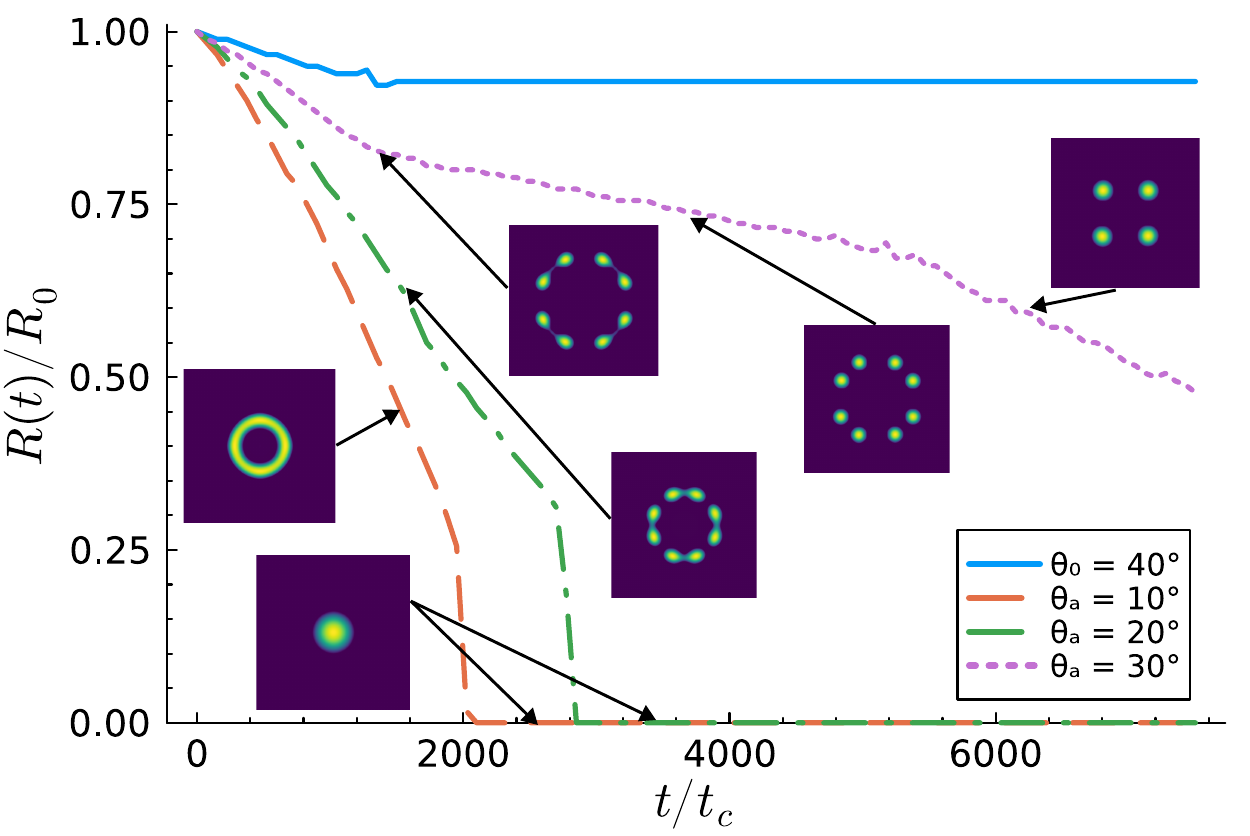}
    \caption{Evolution of the principal radius of a ring-rivulet, $R(t)$, normalized by its initial value $R_0$, for different contact $\theta_a$ at fixed $\theta_b = 40^{\circ}$ (i.e. positive gradient, see Eq.~(\ref{eq:theta_grad})) and $\psi = 0.1$.
    Besides the lines we add heatmap snapshots of the film thickness $h(\mathbf{x},t)$ where yellow indicates a large value and dark blue a small value (colorbars are not scaled).}
    \label{fig:negativewetgrad}
\end{figure}
We consider, now, the linear wettability profile, first with a positive contact angle gradient ($\theta_a < \theta_b=40^{\circ}$ in Eq.~(\ref{eq:theta_grad})), such that the wettability increases towards the centre. 
Fig.~\ref{fig:negativewetgrad} shows measurements of the ring radius $R(t)$ for three different gradients realized with $\theta_a = \{ 10^{\circ}, 20^{\circ}, 30^{\circ} \}$ and on a uniform substrate with contact angle $\theta_0 = 40^{\circ}$, which is reported as a reference. 
The value of the aspect-ratio is set to $\psi_0 = 0.1$ such to be in the range where breakup is expected and hole collapse is inhibited ($\psi_0 \lesssim 0.2$).
In fact, for the uniform case (solid blue line), $R(t)$ stays almost constant for the entire simulation, signalling that the ring ruptures and forms droplets, which essentially do not slide (or slide very little to a new equilibrium radial position) over the substrate.
From the remaining curves, on the contrary, we can see the impact of the wettability gradient that favours ring retraction. 
The decrease of $R(t)$ indicates the occurrence of the wettability-gradient-induced collapse of the inner hole, with steeper gradients corresponding to faster descents.
In particular, for the largest contact angle gradient, retraction is so fast that prevents breakup. 
However, for smaller gradients, unlike the standard collapse process (on uniform substrates), the rivulet breaks up while shrinking, thus forming droplets that eventually re-merge at the centre. 
The substrate modulation is able, then, to determine the coexistence of the two main dewetting paths and allows, at least transiently, the formation of droplets closer to the centre. 
Moreover, the heatmaps of the height field in Fig.~\ref{fig:negativewetgrad} suggest that, tuning the gradient values it is possible to control the evolution of the number of droplets as the ring retracts.\\
\begin{figure}
    \centering
    \includegraphics[width=0.48\textwidth]{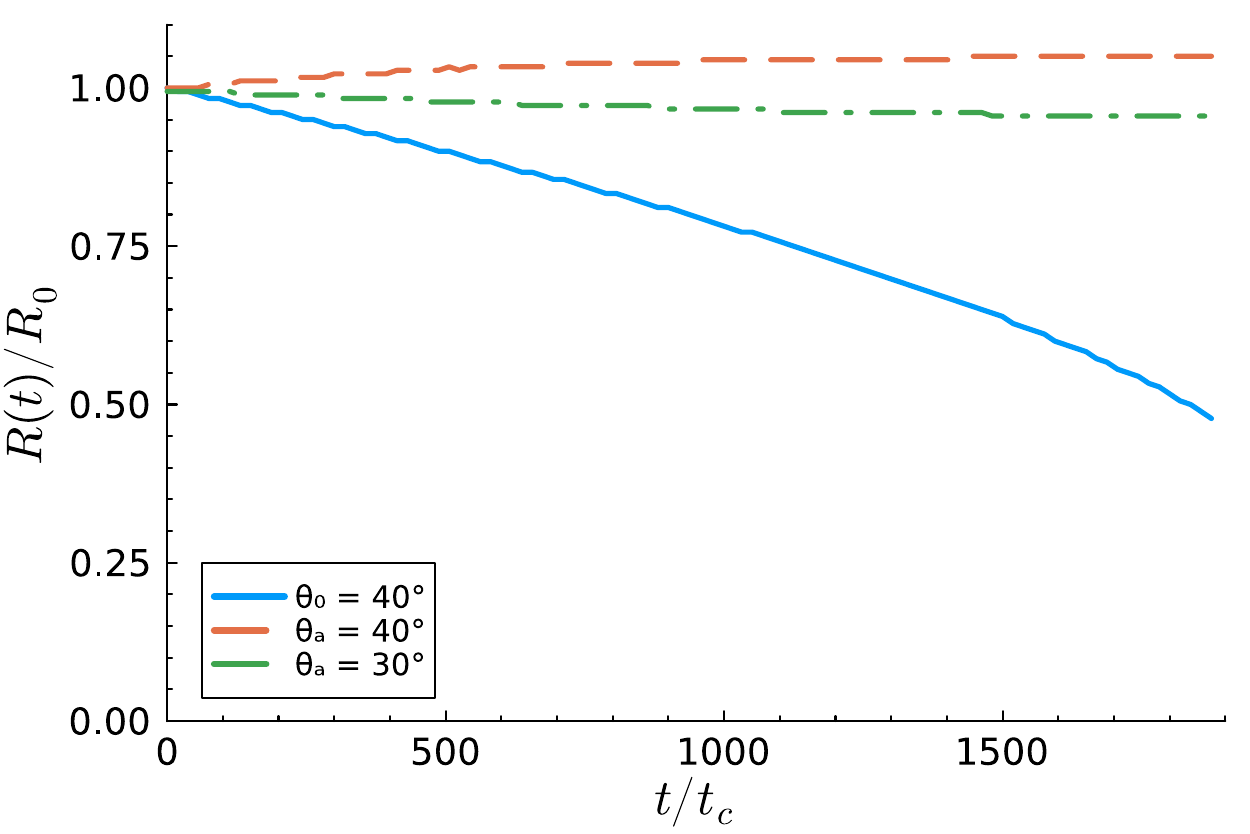}
    \caption{Evolution of the principal radius of a ring-rivulet, $R(t)$, normalized by its initial value $R_0$, for different contact $\theta_a$ at fixed $\theta_b = 40^{\circ}$ (i.e. negative gradient, see Eq.~(\ref{eq:theta_grad})) and $\psi = 0.3$.}
    \label{fig:positivewetgrad}
\end{figure}
We then performed numerical simulations with a negative contact angle gradient ($\theta_a > \theta_b = 20^{\circ}$), thus the substrate is less and less wettable as the centre of the whole is approached.
We set $\psi_0 = 0.3$, so to be in the regime where ring retraction and hole collapse are favoured and the wettability gradient acts as a competing mechanism.
In Fig.~\ref{fig:positivewetgrad} we show $R(t)$ for $\theta_a = \{ 30^{\circ}, 40^{\circ} \}$ and for the uniform case, with $\theta_0 = 40^{\circ}$, which is again reported for comparison.
The monotonic decrease of $R(t)$ on the latter substrate (blue line) indicates, indeed, that the hole is shrinking, as expected for the given value of $\psi_0$. 
Collapse is, instead, inhibited by the outward directed wettability gradient, as the behaviour of the orange and green lines, approaching stationary plateau values, suggest; actually, for the largest gradient, a slight initial increase of the ring radius is even observed. 
This alternative way of removing the collapse mode, then, allows to transiently stabilize rivulets of comparatively large width (with respect to the annular band, for instance), hence to deposit larger volume of liquids in a ring shape of a given hole radius. 

\section{Conclusions}\label{sec:conclu}
We have presented results on the stability and dewetting pathways of liquid ring-shaped rivulets, from mesoscopic numerical simulations of the thin film equation, featuring a disjoining pressure model that allows to consider substrates with various wettability patterns. 
We first considered the dewetting on a uniform substrate, for different values of the contact angle, $\theta_0$. 
In agreement with theoretical predictions, we observed that, for $\psi_0 \approx w/R \ll 1$, the rivulet breaks up into droplets, whose number follows, approximately, the relation $n_d \approx \pi/(2\psi_0)$~\cite{gonzalezStabilityLiquidRing2013}, basically for all contact angles explored.
For $\psi_0 \gtrsim 0.2$, we found that rivulet retraction (hole collapse) occurs, irrespective of the contact angle. 
If the equilibrium state does not seem to be strongly affected by the wettability, the characteristic time scale of breakup and collapse do. 
In the breakup regime ($\psi_0 < 0.2$), the breakup times, $\tau_b$, grow with $\psi_0$ (as expected from models without disjoining pressure~\cite{gonzalezStabilityLiquidRing2013}) only for very small contact angles. 
For $\theta_0 > 10^{\circ}$, $\tau_b$ becomes essentially independent of $\psi_0$. 
In the collapse regime ($\psi_0 > 0.2$), the characteristic time $\tau_c$ is found to decreases with the aspect-ratio as $\tau_c  \sim \psi_0^{-2}$; we provided a phenomenological argument to explain theoretically the origin of such behaviour.\\
We then introduced a wettablity pattern with an annular patch of lower contact angle onto which the ring-rivulet was deposited. For this system, we show that the collapse mode is removed and the rivulet undergoes breakup for any value of the aspect-ratio, though, on time scales comparatively longer (i.e. the pattern makes the rivulet more stable against rupture). 
Moreover, we observed that an increase in contact angle mismatch between patch and background causes the number of formed droplets to deviate from the homogeneous substrate case (and from the theoretical expectation) and become independent of $\psi_0$.\\
Lastly we manage to induce collapse of an otherwise non-shrinking rivulet ($\psi_0<0.2$) by the means of a negative linear radial contact angle profile.
If the sign of this gradient is switched we further achieve a stabilisation of the principle radius for a rivulet that would otherwise collapse ($\psi>0.2$). 
%By means of a linear radial contact angle profile we managed, depending on the sign of the gradient, to either induce hole collapse in an otherwise stable against shrinkage rivulet ($\psi_0<0.2$), or, alternatively, to stabilise to their initial principal radius rivulets with aspect-ratios that would lead to collapse ($\psi>0.2$). \\
A perspective venue for future development of this work could be to introduce time-dependent wettability patterns, as done in Zitz et al.~\cite{zitzControllingDewettingMorphologies2023}, which could provide a new handle to allow better control of the instabilities, and possibly opening the way to new unexpected dewetting pathways.

\section*{Author Contributions}
S. Z, A. S. and J. R.  contributed equally to the design and conceptualization. S. Z. carried out the numerical experiments. S. Z. and A. S. performed the analysis of the results and wrote the original draft. 

\section*{Data availability}
The data is readily available at \href{https://github.com/Zitzeronion/Ring_rivulets}{https://github.com/Zitzeronion/Ring\_rivulets}

\section*{Conflicts of interest}
There are no conflicts to declare.

\section*{Acknowledgements}
S. Z. and J. R. acknowledge the financial support from the Independent Research Fund Denmark through a DFF Sapere Aude Research Leader grant (grant number 9063-00018B).

%%%END OF MAIN TEXT%%%


\balance


%%%REFERENCES%%%
%\bibliography{rsc} %You need to replace "rsc" on this line with the name of your .bib file
% \bibliographystyle{rsc} %the RSC's .bst file
\providecommand*{\mcitethebibliography}{\thebibliography}
\csname @ifundefined\endcsname{endmcitethebibliography}
{\let\endmcitethebibliography\endthebibliography}{}

\appendix
\section{Numerical model and parameters}\label{app:numerics}
The numerical model is based on a single relaxation (SRT) time lattice Boltzmann method. 
The relaxation time $\tau$ is set to $\tau = 1$ in all our numerical experiments. 
This means that the fluid's kinematic viscosity $\nu$ is set to $\nu = 1/6$ and kept constant for all simulations.  

The surface tension $\gamma$ is set to $\gamma = 10^{-2}$ for all results presented here.
We did in fact simulate with different values of $\gamma$ but found that the effect is an overall rescaling of the time scale, larger $\gamma$ speeds up all dynamics, lower $\gamma$ values slows them down.

Lastly $\alpha_{\delta}(h)$ is a substrate friction term that mimics a slip boundary condition with an effective slip length $\delta$
\begin{equation}\label{eq:alphafric_app}
\alpha_{\delta}(h) = \frac{6h}{(2 h^2 + 6 \delta h + 3 \delta^2)}.
\end{equation}
By introducing a precursor layer $h_{\ast}$ and a slip length $\delta$ we regularize the contact line divergence~\cite{huhHydrodynamicModelSteady1971}. 
The slip length lies within the weak/intermediate slip regime~\cite{peschkaSignaturesSlipDewetting2019,fetzerQuantifyingHydrodynamicSlip2007, munchLubricationModelsSmall2005} and the thickness of the precursor film is set to $h_{\ast} = 0.05$.

By applying this solution for $\mathbf{u}$ to the continuity equation we have
\begin{equation}\label{eq:thinsolve_app}
     \partial_t h(\mathbf{x},t) = \nabla\cdot\left(M_{\delta}(h)\nabla p_{\mbox{\tiny{film}}}\right),
\end{equation}
with the mobility function $M_{\delta}(h) = \frac{h^2}{\mu\alpha_{\delta}(h)}$ which for the no-slip boundary condition $(\delta \rightarrow 0)$ reduces to $M_{0}(h) = h^3/3\mu$.
Without loss of generality we set $\rho_0 = 1$ and thus the dynamic viscosity is $\mu = \rho_0 \nu = 1/6$. 

The numerical domain consists of a square lattice with $512\Delta x$ in both horizontal directions.
We further use biperiodic boundary conditions at the edges of the domain. 

\section{Derivation of equation (\ref{eq:modelC4})}\label{sec:derivation}
Let $\theta_A$ and $\theta_B$ be the contact angles at the advancing and receding contact lines, respectively, we have, by the Cox-Voinov law, that 
\begin{equation}\label{eq:modelCV}
\theta_A^3 - \theta_R^3 \propto Ca = \frac{\mu}{\gamma} U = - \frac{\mu}{\gamma} \dot{R}(t)
\end{equation}
(the minus sign comes from the fact that the ring radius is decreasing), whence, if we write 
$\theta_A = \theta_0 + \delta \theta$ and $\theta_R = \theta_0 - \delta \theta$ 
(with $\delta \theta \ll \theta_{A,R}$), we get
\begin{equation}\label{eq:modelC2}
- \frac{\mu}{\gamma} \dot{R}(t) \approx 6 \theta_0^2 \delta \theta.
\end{equation}
Moreover we know from the static solution for the shape of the ring rivulet that, for small angles (i.e. such that one can approximate $\tan \theta \approx \theta$) $\theta_o \approx f(\psi_0) \theta_i$ ($\theta{i,o}$ being the contact angles at the inner and outer contact lines, respectively), where $f(x)$ is a function that, for small argument, goes as $f(x) \sim 1-x$~\cite{gonzalezStabilityLiquidRing2013}.
Assuming this relation valid at any time, we can write, $\theta_R \approx (1-\psi(t))\theta_A$, or also $\delta \theta \approx \psi \theta_0/2$, which, plugged into (\ref{eq:modelC2}) gives
\begin{equation}\label{eq:modelC3}
- \frac{\mu}{\gamma} \dot{R}(t) \approx 3 \theta_0^3 \psi.
\end{equation}
Inserting the expression for the characteristic time $t_c = \mu r_0/\gamma$ and the (instantaneous) aspect-ratio $\psi(t) = 2r_0\sin \theta_0/R(t) \approx 2 r_0\theta_0/R(t)$ in (\ref{eq:modelC3}), 
we obtain equation (\ref{eq:modelC4})
\begin{equation}
-R\dot{R} = - 2\frac{d R^2}{d t} \approx 6 \, \theta_0^4 \, \frac{\gamma r_0}{\mu} = 6 \, \theta_0^4 \, \frac{r_0^2}{t_c}.
\end{equation}

\end{document}